\documentclass{JHEP3}
\preprint{MAN/HEP/2007/6}

\usepackage{cite}
\usepackage{epsfig}
\usepackage{color}
\let\normalcolor\relax
\usepackage{graphics}
\usepackage{inputenc}
\usepackage{xspace}
\inputencoding{latin1}
 \renewcommand\email[1]{{\scriptsize\tt\href{mailto:#1}{#1}}}
\newcommand{\gtaet}{\raisebox{-0.8mm}
{\hspace{1mm}$\stackrel{>}{\sim}$\hspace{1mm}}}
\newcommand{\ltaeq}{\raisebox{-0.8mm}
{\hspace{1mm}$\stackrel{<}{\sim}$\hspace{1mm}}}

\newcommand{\M}{\ensuremath{\mathbf{M}}}
\newcommand{\G}{\ensuremath{\mathbf{\Gamma}}}
\newcommand{\Sv}{\ensuremath{\mathbf{S}}}
\newcommand{\ubar}{\ensuremath{\overline{u}}}

\newcommand{\as}{\ensuremath{{\alpha}_{s}}}

\newcommand{\aem}{\ensuremath{{\alpha}_{EM}}}
\newcommand{\kT}{\ensuremath{\vec{k}_{\perp}}}
\newcommand{\kTone}{\ensuremath{\vec{k}_{1\perp}}}
\newcommand{\kTtwo}{\ensuremath{\vec{k}_{2\perp}}}

\newcommand{\shat}{\ensuremath{\hat{s}}}
\newcommand{\pidlla}{$\pi^2$DLLA }
\newcommand{\lla}{full LLA }

\def\beq{\begin{equation}}
\def\eeq{\end{equation}}
\def\beqa{\begin{eqnarray}}
\def\eeqa{\end{eqnarray}}

\def\e{r}

\newcommand{\eqref}[1]{Eq.~(\ref{#1})\xspace}
\newcommand{\eqrefs}[1]{Eqs.~(\ref{#1})\xspace}

\newcommand{\figref}[1]{Fig.~\ref{#1}}

\skip\footins = 1\bigskipamount plus 2pt minus 4pt                              

\title{\boldmath Soft gluons in Higgs plus two jet production}

\author{Jeff Forshaw and Malin Sjödahl\\
School of Physics \& Astronomy, University of Manchester, \\
  Oxford Road, Manchester M13 9PL, U.K.\\
  E-mail: \email{jeff.forshaw@manchester.ac.uk}
    and \email{malin.sjodahl@manchester.ac.uk}}
  
  \abstract{We investigate the effects of an all order QCD resummation 
    of soft gluon emissions for Higgs boson production in association with two hard
     jets. We consider both the gluon-gluon fusion and weak boson fusion processes
     and show how to resum a large part of the leading logarithms in the jet veto scale.
     Our resummation improves on previous analyses which also aim to include
     the effects of multiple soft gluon radiation. In addition we calculate the interference between weak
     boson fusion and gluon-gluon fusion and find that it is small.
  }

\begin{document}
 
\sloppy
 
\section{Introduction}
\label{sec:intro}

Searching for the Higgs boson is one of the main tasks of the LHC and its production in association with two low-angle (tag) jets should provide an effective way of reducing backgrounds. In the literature, most of the attention has been focussed upon the contribution from weak boson fusion (WBF), see \figref{fig:LO}(a), although the Higgs can also be produced through gluon fusion, as illustrated in \figref{fig:LO}(b). Detecting the Higgs boson in association with tag jets should provide the opportunity to gain important information on the coupling of the Higgs to weak bosons \cite{Zeppenfeld:2000td,Plehn:2001nj,Belyaev:2002ua,Duhrssen:2004uu,Hankele:2006ma,Berger:2004pc} and even to the top quark \cite{Klamke:2007cu}.

To obtain these couplings separately one would like experimentally to differentiate the two cases. Fortunately there are several differences which arise. In WBF, the two highest $k_{\perp}$ jets (which we define to be the tag jets) tend also to be the high rapidity quark jets which are responsible for radiating the fusing weak gauge bosons. This is to be contrasted with the case of gluon fusion where the highest $k_{\perp}$ jets need not be energetic low-angle quark jets. As a result, the tag jets in gluon fusion tend to be much less forwardly peaked and form a much lower dijet invariant mass than the tag jets present in WBF.  
The dependence upon the azimuthal angle between the tag jets is also sensitive to the difference between WBF and gluon fusion. In particular, the tag jets in the WBF channel are approximately independent of the azimuthal angle whereas in the gluon fusion channel they have a $\cos^2\phi$ dependence (for a CP even Standard Model Higgs).

\FIGURE[t]{\epsfig{file=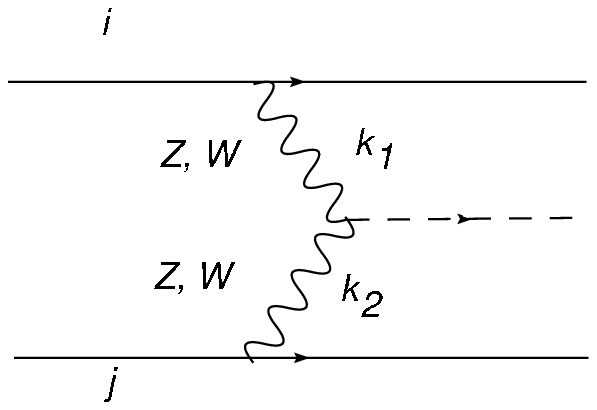,width=5.5cm}\,\,\,\,\epsfig{
    file=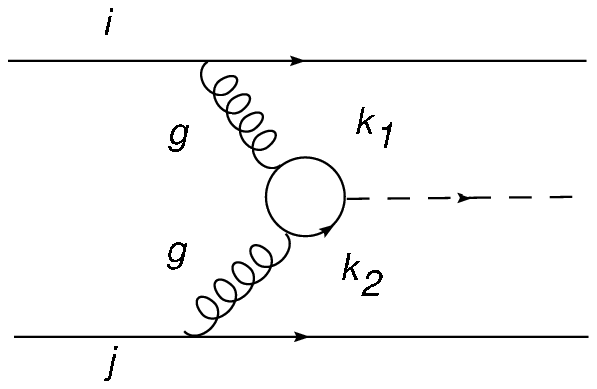,width=5.5cm}
  \caption{\label{fig:LO} The lowest order $q+q \rightarrow H +q +q$
    diagrams in the eikonal limit,
    (a) for weak boson fusion;
    (b) for gluon-gluon fusion with the Higgs being produced 
    via a top quark loop.}}

In this paper, we will focus on another important difference, namely the tendency for the weak (color singlet) exchange to radiate fewer particles into the region between the tag jets. Indeed, this difference is often exploited in order to produce a WBF rich sample of events; typically one insists that there be no additional jets in between the tag jets with the transverse momentum above some value, $Q_0 \sim 20$ GeV. The tendency of WBF to radiate less has been studied previously in the context of Higgs plus two-jet production in association with the emission of one extra parton \cite{Khoze:2003af,DelDuca:2004wt}, by employing parton showering from event generators \cite{DelDuca:2006hk}, by using the truncated shower approximation (TSA) \cite{PhysRevLett.55.2752,Rainwater:1998kj,PhysRevD.60.113004,Barger:1994zq,Duff:1993ut,PhysRevD.54.6680} and via calculations based on the soft gluon exponentiation approach
\cite{Barger:1994zq,PhysRevD.54.6680}. Fixed order calculations \cite{Khoze:2003af,DelDuca:2004wt,Berger:2004pc,Campbell:2006xx} inherently suffer from the problem that as soon as  $\sigma^{\mathrm{3 jet +H}}  \gtaet \sigma^{\mathrm{2 jet +H}}$ then an all order resummation is desirable. In this paper we shall show how to improve fixed order calculations by summing to all orders in the strong coupling what is likely to be the dominant subset of the leading logarithms in the ratio of the hard interaction scale and the jet veto scale $Q_0$. Our calculations ought to improve upon the existing methods when it comes to accounting for multiple gluon emission in between the tag jets. Before we present our calculations we shall briefly review the aforementioned approaches.

In the truncated shower approximation  
\cite{PhysRevLett.55.2752,Rainwater:1998kj,PhysRevD.60.113004,Barger:1994zq,Duff:1993ut,PhysRevD.54.6680}, fixed order QCD is used to compute the cross-section for the real emission of one additional parton, i.e. $Hjjj$. Two of the partons are the tag jets and the presence of the third (mini-)jet, of transverse momentum $p_T$, permits a study of the role of a jet veto in the region between the tag jets. However, integrating the three parton cross-section over the phase-space of the third (mini-)jet will lead to a divergent cross-section. In the TSA, this divergence is regulated by modifying the three parton cross-section ($\sigma_1$)  to
\begin{equation}
d\sigma_1^{TSA}    =  (1- e^{-p_T^2/p_{TSA}^2})d\sigma_1.
\end{equation}
The scale $p_{TSA}$ is chosen so that the integrated cross-section is equal to the two parton cross-section, i.e.
\begin{equation}
\int_0^{\infty} dp_T \; \frac{d\sigma_1^{TSA}}{dp_T} = \sigma_0
\end{equation}
where $\sigma_0$ is the lowest order cross-section (either for WBF or gluon fusion).
This is clearly a rather crude phenomenological model. It has the virtue that it contains the complete information of the fixed order calculation for single parton emission however it will fail for small enough veto scales, $Q_0 \ltaeq p_{TSA}$, whereupon multiple parton emission starts to become important. The enhanced probability for radiation in gluon fusion events leads to a larger $p_{TSA}$ than in the case of WBF.

The soft gluon exponentiation model \cite{Barger:1994zq,PhysRevD.54.6680} provides for a more sophisticated treatment in the case of multiple emissions by exploiting the fact that the leading behaviour comes from the emission of soft gluons which are assumed to exponentiate. The cross-section for emitting no gluons with transverse momenta above $Q_0$ is given by
\begin{equation}
  d\sigma =\exp\left(-\frac{1}{\sigma_{0}}
  \int_{Q_0}^{}dq_T \frac{d \sigma_{1}}{dq_T} \right) d\sigma_0
\end{equation}  
where $\sigma_{1}$ is the cross-section for emitting one extra gluon into the veto region. As we shall see, this approach does contain an important part of the leading logarithmic soft gluon contribution, although it does not include any colour mixing.

Like the soft gluon exponentiation model, parton shower algorithms (as implemented in the all-purpose event generators \cite{Sjostrand:2006za,Corcella:2002jc}) succeed in systematically improving beyond fixed order by including a major part of the colour diagonal soft gluon evolution. However, the parton showers  also do not include any colour mixing contributions and the soft gluon evolution is usually performed in the large $N$ approximation\footnote{$N$ is the number of colours}. We should add that event generators have the virtue that they deal with hadronization, which is of course not possible in a purely perturbative treatment.  Hadronization is, however, expected to be ``gentle'' in the sense that hard enough jets are not much affected by it. Nevertheless, the role of multiple parton interactions and the soft underlying event do lead to important features of the hadronic final state which are inaccessible to perturbative QCD. It is common to chose the veto scale $Q_0$ to be sufficiently large (e.g. $\gtaet 20$ GeV) so that sensitivity to the underlying event is reduced \cite{Asai:2004ws,Butterworth:2002tt}. We shall not consider the underlying event any further in this paper but its impact clearly needs to be accounted for when it comes to comparing to data. We shall also assume that the Higgs boson is sufficiently long lived that we do not need to consider the complexities which arise upon its decay to coloured particles.

An alternative, experimental based, approach which should help us to understand the radiation
pattern in Higgs production has been proposed in \cite{PhysRevD.54.6680,Khoze:2002fa}.
Here it is suggested to probe the mini-jet distribution in hard color singlet scattering events using measurements of the process $q q \to q q + Z $ via $W$-fusion.

Soft, wide angle gluon resummations similar to those we perform here (but without Higgs production) have been 
presented in \cite{Kidonakis:1998nf,Oderda:1998en,Oderda:1999kr,Berger:2001ns,Berger:2002ig,Appleby:2003sj}. Generically, one is interested in restricting radiation from a hard scattering at scale $Q$ so that, in some specified region of the final state phase-space, there should be no radiation above the veto scale $Q_0$. The goal is at present limited to summing the leading logarithms, i.e. terms of the form
 \begin{equation}
\sim \alpha_s^n \log(Q/Q_0)^n.
\end{equation}
We shall usually find it convenient to display our results in terms of the evolution variable
\begin{equation}
  \label{eq:xi}
  \xi \equiv \int_{Q_0}^Q \as(q_T) \frac{dq_T}{q_T}.
\end{equation}
At the LHC a typical $Q_0$ is expected to be $20$ GeV. Assuming a hard scale of $200$ GeV then gives $\xi\sim 0.3$ with $n_f=5$ and $\as(m_Z)=0.12$. If $Q_0$ can be pushed down to $5$ GeV,  $\xi$ could reach well above $0.5$. Recall that the larger one can make $\xi$ the more dominant one expects the WBF contribution to be.

In this paper we will make an all order resummation of a part of this leading logarithmic series. We will, however, not include those leading logarithms which arise by virtue of the fact that the process we consider is ``non-global"  \cite{Dasgupta:2001sh,Dasgupta:2002bw}.  The problem of summing the leading non-global logarithms is a thorny one since the colour structure makes anything other than a leading $N$ approximation impossible at the present time. Fortunately for Higgs production, the impact of the non-global logarithms is expected to be small provided we do not choose a particularly small cone radius \cite{Appleby:2002ke,Delenda:2006nf}. For example, in \cite{Delenda:2006nf} the non-global logarithms were estimated in the two-quark case (e.g. $e^+e^- \to q \bar{q}$) and, for a cone radius $R=1$ and evolution variable $\xi = 0.9$, found to be a 5\% correction to the primary emission result. For smaller $\xi$ values the effect diminishes still further since the non-global logarithms enter first at order $\xi^2$. Apart from the non-global logarithms, we shall also neglect a leading contribution to the primary emission contribution which arises as a result of implementing a jet algorithm \cite{Banfi:2005gj,Delenda:2006nf}. This too is a $\xi^2$ effect and therefore is smaller at lower values of $\xi$. The results quoted in \cite{Delenda:2006nf} indicate that these logarithms begin to play a role for relatively large $\xi$ values, i.e. $\xi \gtaet 0.5$ for $R=1$ (diminishing as $R^3$ for smaller $R$). Following \cite{Delenda:2006nf}, these jet algorithm dependent corrections can be systematically computed and so the results we present here could be further improved upon if need be.
Finally, we shall ignore the potential impact of the ``super-leading logarithms'' discussed in \cite{Forshaw:2006fk}. At the present time, the existence of these logarithms is uncertain. In any case, one might expect any such contribution to the Higgs plus jet production process to be modest since they enter first at order $\as^4$ and are subleading in $N$.

We will of course re-confirm that the radiation pattern is very different for WBF and gluon-gluon fusion. We will also show that the LHC will explore the kinematic region where higher order soft gluon corrections are important and fixed order calculations are inadequate. As for the interference (between gluon fusion and WBF contributions), this turns out to be surprisingly small. This is in part a result of the soft gluon evolution, but it is also a consequence of a tree level helicity dependent cancellation present in the eikonal approximation. The source of interference we consider is quite distinct from the crossed-channel interference discussed in \cite{Andersen:2006ag}, which vanishes in the eikonal approximation.

The outline of this paper is as follows. First in Section \ref{sec:LO} we have a quick look at the lowest order cross-sections to set the scene for what follows. Then, in Section \ref{sec:GR}, we explain the relevant theory and show how to accommodate the extra outgoing Higgs boson. We also briefly discuss the recent two-loop calculations of \cite{MertAybat:2006mz}. Numerical results are then presented and commented upon in Section \ref{sec:Num}. Finally, we draw our conclusions in Section \ref{sec:Conclusions}.

\section{Lowest order}
\label{sec:LO}

For both electroweak $ZZ$ and $WW$ fusion processes, shown in \figref{fig:LO}a, and the QCD gluon-gluon fusion process, shown in \figref{fig:LO}b, we shall compute the lowest order cross-section in the eikonal approximation. This is a good approximation in the case where the final state jets are at low angles \cite{DelDuca:2002bf}. In any case, our goal in this paper is not to provide precise predictions for the LHC rather it is to show how to include the leading effects of QCD radiation. In the eikonal approximation, only the $t$-channel contributions illustrated in \figref{fig:LO} are important; the vertex factor $\ubar \gamma_\mu u$ simplifies to  $\ubar \gamma_\mu u \propto 2 p_\mu$
and the parton helicity goes unchanged. In the electroweak vertices the factor $\gamma^5$  results in an extra minus sign in the case of left handed fermions, i.e. $\ubar_L \gamma_\mu \gamma^5 u_L \propto -2 p_\mu$. 

Using the notation of \figref{fig:LO} and the conventions of \cite{PS}  the WBF amplitudes are then
\begin{eqnarray}
  \label{eq:EWBorn}
  M_{ZZ}&=&\frac{16 \pi \aem \shat \; m_Z^2 f_i f_j}
  {(m_Z^2+\kTone^2)(m_Z^2+\kTtwo^2) \; v \cos^2\theta_W \sin^2\theta_W},
  \nonumber \\
  M_{WW}&=&\frac{8 \pi \aem \shat \; m_W^2}
  {(m_W^2+\kTone^2)(m_W^2+\kTtwo^2) \; v \sin^2\theta_W}
\end{eqnarray}
with  $v$ being the Higgs boson vacuum expectation value, $v^2=(G_F\sqrt{2})^{-1}$.
Here it is to be understood that the charged current amplitude vanishes for right-handed 
quarks. In principle the amplitude should contain CKM matrix elements  but since we are not interested in the flavor of the outgoing quarks we neglect them.  In the case of $Z$ boson fusion the factors 
\begin{equation}
f_i=T^3_{Li} -Q_i \sin^2\theta_W
\end{equation}
account for the helicity/flavour dependent  factor from the $Z$-vertex (with $T^3_{Li}=0$ for right-handed quarks).

For the gluon fusion process we work in the $m_t \rightarrow \infty$ limit 
\cite{Dawson:1990zj,Shifman:1979eb,Djouadi:1990aj} and note that the approximation is good at the level of a few percent even for $m_H=m_t$ \cite{DelDuca:2001fn}.
It  has also been shown to be a good approximation even in the limit where the dijet invariant mass is large provided $k_{\perp} \ltaeq m_t$ \cite{DelDuca:2001fn}. Up to a colour factor, the amplitude is (for quark or gluon external states)
\begin{equation}
  \label{eq:QCDBorn}
  M_{gg}=\frac{8\as^2 \shat \; \kTone \cdot \kTtwo }{3 \kTone^2 \kTtwo^2 v}.
\end{equation}
Comparing the WBF and gluon fusion results (\eqrefs{eq:EWBorn} and \eqref{eq:QCDBorn})
we note that gluon fusion dominates for small enough or large enough tag jet transverse momenta.  At the energy scales interesting for LHC measurements there is, however, a very important intermediate $\kT$ range where the WBF process dominates. We also note that $|M_{gg}|^2$ depends on the azimuthal angle as $\cos^2\phi$ whereas  $|M_{ZZ}|^2$ and $|M_{WW}|^2$ are flat in azimuth. 

\section{Gluon resummation}
\label{sec:GR}

To compute the corrections to the lowest order amplitudes of the previous section arising from QCD radiation we follow closely the calculation of the ``gaps between jets" process presented in \cite{Kidonakis:1998nf,Oderda:1998en,Oderda:1999kr,Berger:2001ns,Berger:2002ig,Appleby:2003sj}. This process has been widely studied in the literature \cite{Oderda:1998en,Oderda:1999kr,Appleby:2003sj,Mueller:1992pe,Cox:1999dw,Motyka:2001zh,Enberg:2001ev,Forshaw:2005sx} and in experiments at HERA \cite{Adloff:2002em,Derrick:1995pb} and the Tevatron \cite{Abbott:1998jb,Abe:1997ie}. In the gaps between jets process,  one is interested in final states which contain two hard jets at a scale $Q$ with the restriction that there be a region (of size $Y$ in rapidity) between the jets into which no further jets with $p_T > Q_0$ can be emitted. This situation is very similar to the Higgs plus two jets process. The only difference is the presence of a third colour neutral particle (the Higgs) in the final state and not surprisingly we can carry over much of the formalism.

In order to account for the presence of a Higgs, we follow the calculation presented for the five parton evolution in \cite{Kyrieleis:2005dt} in conjunction with the result of \cite{Oderda:1998en,Oderda:1999kr}. In this section, we focus upon quark-quark scattering. This is the whole story for the WBF process but clearly the gluon fusion process can be generated using either quarks, anti-quarks or gluons within the colliding protons. The generalization to processes initiated by anti-quarks or gluons is straightforward and we present
final results for all of the relevant subprocesses. In the $qq \to qqH$ case, the final result is
\begin{equation}
  \label{eq:M}
  \M=\exp\left(  -\frac{2}{\pi}
	 {\displaystyle\int\limits_{Q_{0}}^{Q}}
	 \alpha_{s}(q_T) \frac{dq_{T}}{q_{T}} \G \right)  \M_{0},
\end{equation}
where
\begin{equation}
  \label{eq:Gamma}
  \G = \frac{C_F}{2}\rho(Y,y_3,y_4) \mathbf{1}+
  \left(
  \begin{array}
    [c]{cc}
    0  & \frac{N^{2}-1}{4N^{2}}i\pi\\
    i\pi & (\frac{N}{2}Y-\frac{1}{N}i\pi)
  \end{array}
  \right)
\end{equation}
and
\begin{equation}
  \label{eq:rho}
  \rho(Y,y_3,y_4)=\frac{1}{2}\left( \log\frac{\sinh(|y_3|+Y/2)}{\sinh(|y_3|-Y/2)}+ 
  \log\frac{\sinh(|y_4|+Y/2)}{\sinh(|y_4|-Y/2)}\right)-Y. 
\end{equation}
In the frame in which the incoming quarks collide head on with equal energies, the rapidities of the outgoing quarks jets are $y_3$ and $y_4$ and $Y (< |y_3-y_4|-2R)$ is the region in rapidity between the two tag jets into which radiation above $Q_0$ is forbidden ($R$ is the radius parameter). The effect of including the Higgs boson is essentially kinematical since, unlike the gaps between jets case, we do not require that $y_3 = -y_4$. Numerically the effect is small in the eikonal region since $\rho$ is a small constant for large enough $Y$ and $R$ not much smaller than 1.

We need also to explain the colour basis implicit in the above. We use a singlet-octet basis such that the lowest 
order amplitude for neutral current quark scattering is
\begin{equation}
\M_0 =   \left(
  \begin{array}
    [c]{c}
    M_{ZZ}\\
    M_{gg}
  \end{array} \right)\;.
\end{equation}
To reconstruct the cross-section, we must restore the colour factor for the gluon fusion process, i.e. we need
$\sigma=\M^{\dagger} \mathbf{S} \M$ where
\begin{equation}
  \label{eq:S}
  \mathbf{S}=
  \left(
  \begin{array}
    [c]{cc}
    1 & 0\\
    0 & \frac{N^2-1}{4N^2}
  \end{array}
  \right).
 \end{equation}
The amplitude for charged current fusion, $M_{WW}$, does not interfere with the neutral current fusion processes and soft gluon evolution therefore proceeds as in \eqref{eq:M} with $\M_0^T = (M_{WW},0)$. The amplitudes for $q\bar{q} \to q\bar{q}H$,
$qg \to qgH$ and $gg \to ggH$ are obtained analogously and the relevant evolution matrices are presented in the appendix.

It is now interesting to reconsider the various radiation models discussed in the introduction and look at them in the light of Eqs. (\ref{eq:M})--(\ref{eq:rho}). Clearly the truncated shower approximation has no direct correspondence in terms of a $\G$. Exponentiating the differential probability for emitting an extra gluon, as is done in the soft gluon exponentiation model, should catch the diagonal $C_F$ dependent parts of $\G$, as well as the $Y$ dependent part in the octet-octet entry. The imaginary parts, coming from Coulomb gluon exchanges which generate on shell intermediate quarks, will however not appear when only real emissions are considered and the octet-singlet mixing parts of $\G$ are thereby lost. 
In Monte Carlo parton showers we anticipate that the leading $N$ and diagonal $C_F$ dependent terms are included. Once again the Coulomb gluon $i \pi$ terms are not included. 

Let us make a few general remarks about the soft gluon evolution. Firstly, the Coulomb gluon contributions which generate the $i \pi$ terms are just the exchanges which in QED would generate an overall phase in the amplitude whereupon they would have no effect on observables. In the non-abelian QCD case they do not, however, simply build a phase and they play a potentially crucial role. In particular, it is through these terms that the singlet and octet sectors are able to communicate with each other. Indeed Coulomb gluon exchange is in this way seen to be responsible for generating the QCD pomeron. 

For large enough $Y$ we may neglect the $C_F \rho$ dependent terms in $\Gamma$ and we refer to this approximation as the $\pi^2$DLLA. In this approximation, the lowest order in $\as$ where soft gluon effects enter into WBF contribution to the cross-section is ${\cal{O}}(\as^3)$. This is not the case for octet exchange, where the leading contribution at large $Y$ is ${\cal{O}}(\as NY)$. Thus we expect that the $C_F \rho$ dependent terms should be relatively more important in WBF than in gluon fusion. Note also that the interference between WBF and gluon fusion starts at ${\cal{O}}(\as^2)$ in the cross-section.
 
Recently Mert Aybat, Dixon and Sterman \cite{MertAybat:2006mz} have confirmed that the two-loop soft gluon evolution matrix  is proportional to the one-loop matrix $\Gamma$ and hence that its effect can be accounted for if one uses the strong coupling in the CMW scheme \cite{Catani:1990rr}, i.e. our calculations can be extended to include their result via a re-definition of $\xi$:
\begin{equation}
  \label{eq:xiEff}
  \xi \to \int_{Q_0}^Q \as(q_T) \left( 1 + K \frac{\as(q_T)}{2 \pi} \right) \frac{dq_T}{q_T}
\end{equation}
with 
\begin{equation}
  \label{:eq:K}
  K=N \left(\frac{67}{18}-\frac{\pi^2}{6} -\frac{5}{9} \; n_f\right).
\end{equation}
The effect is thus a small downward shift in $\xi$. Although the two-loop evolution matrix is an important component of the higher order calculation it is not the whole story and so at present a full next-to-leading logarithmic calculation is not possible.

\section{Numerical results}
\label{sec:Num}

In \figref{fig:RelAmp2} we show the square of the helicity averaged lowest order amplitudes,
\eqref{eq:EWBorn} and \eqref{eq:QCDBorn}, in the case of a Higgs produced with zero transverse momentum, i.e. 
$\kTone=-\kTtwo$. We have used a one-loop running $\as$ with $n_f=5$, $\as(m_Z=91.2\; \mbox{GeV})=0.120$,
a constant $\aem$ of 1/127.9 and $\sin^2\theta_W=0.231$. For $ZZ$ fusion we have chosen to 
display the squared amplitude for u-type quarks only, but clearly the result for other partonic combinations can be obtained by a simple re-scaling of the displayed curve.

\FIGURE[h]{\epsfig{file=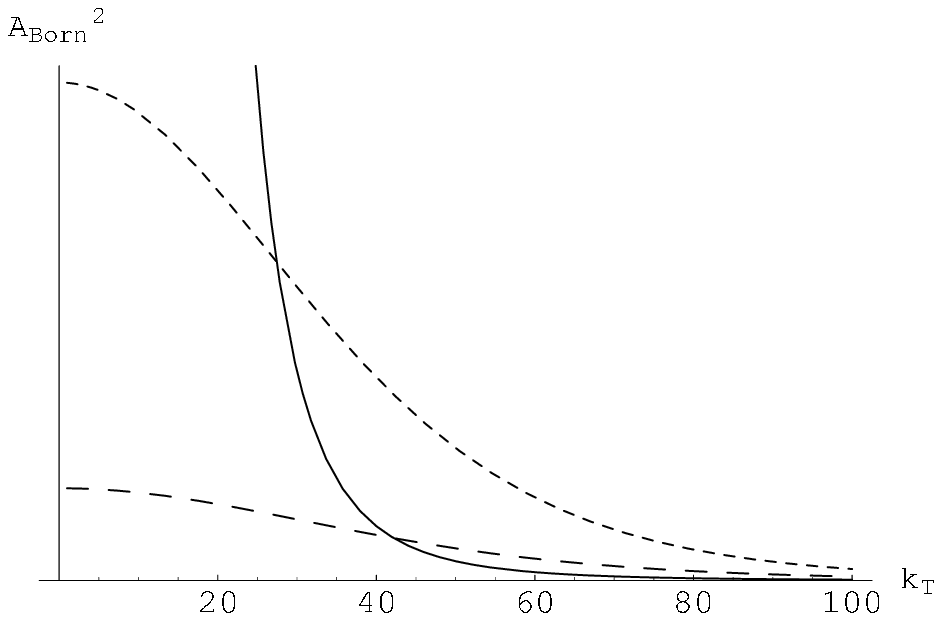,width=8cm}
  \caption{\label{fig:RelAmp2} The relative squared amplitudes of 
    \eqref{eq:EWBorn} and \eqref{eq:QCDBorn}. The solid line is gluon-gluon fusion, the long dashed line $ZZ$ fusion and the short dashed line WW fusion.}}

In order to obtain realistic hadron level cross-sections the parton distribution functions must be folded in. However we do not perform that step here, preferring instead to concentrate on the physics associated with soft gluon radiation. Saving the interference for later, we now move on to consider the QCD evolution of the gluon-gluon fusion and WBF amplitudes.

The various squared amplitudes, divided by the corresponding lowest order results, are displayed in \figref{fig:Evolution} as functions of $\xi$ for various values of $Y$ and for $|y_3|=|y_4|=Y/2+R$ with a cone radius $R=1$. To get a handle on the \lla as compared to the $\pi^2$DLLA, the \lla evolution is displayed with solid lines and the \pidlla with dashed.  We can see clearly the effect of the larger emission probability for the gluon fusion processes and that the $\pi^2$DLLA typically provides a good 
approximation. Perhaps most striking is the fact that the suppression with increasing $\xi$ is much the same for all of
the gluon fusion processes, i.e. it is not very much dependent upon whether the primary jets are quarks or gluons\footnote{Of course the radiation in $qq$ and $q\bar{q}$ processes is very similar.}. This is
to be expected and is a result of colour coherence. Large angle soft gluon emission is not sensitive to the colour charge of
the external partons, rather it is as if the radiation is from the exchanged $t$-channel gluons. In fact, in the large $N$
approximation, the suppression is exactly equal for all partonic subprocesses as a result of the diagonal nature of the evolution
matrices (listed in the appendix), i.e. 
$$ \frac{\sigma}{\sigma_0} \sim \exp \left( -\frac{2NY\xi}{\pi}\right).$$
Collinear gluon emission is a different matter however, and its effect appears through the diagonal evolution proportional to
$\rho$. In that case, radiation off gluon and quark jets occurs in proportion to their respective colour charges. For large enough cone radii, the gap is sufficiently removed from the direction of the outgoing partons that collinear evolution
is unimportant (i.e. $\rho$ is small). Hence the relatively small difference between the solid and dashed curves in all of the panes
in \figref{fig:Evolution}. In order to enhance the differences between quark and gluon jets, one would need to take a small
cone radius in order to enhance the role of $\rho \sim \log(1/R)$ at small $R$.

We note that the WBF cross-section (the uppermost curves in the first row of the figure) actually rises slightly as we go from $Y=3$ to $Y=6$.  This is a result of the fact that for singlet exchange leading logarithmic soft gluon effects are contained entirely in the $\exp(-\frac{C_F}{\pi} \xi \rho)$ term at large $Y$ whilst at lower values of $Y$ the mixing into octet and the associated suppression plays a role. Thus at large $Y$ we see the dominance of pure singlet evolution in the WBF process. This result is not so transparent in the octet-singlet basis we use here but it is apparent in a basis in which the evolution is diagonal \cite{Forshaw:2005sx}.  We caution however that the $\as^n Y^n$ terms will be important for large $Y$ and (apart from the double logarithmic $\as^n Y^n \log^n(Q/Q_0)$ subset) they are not included in the approach presented here.

\FIGURE[t]{\epsfig{file=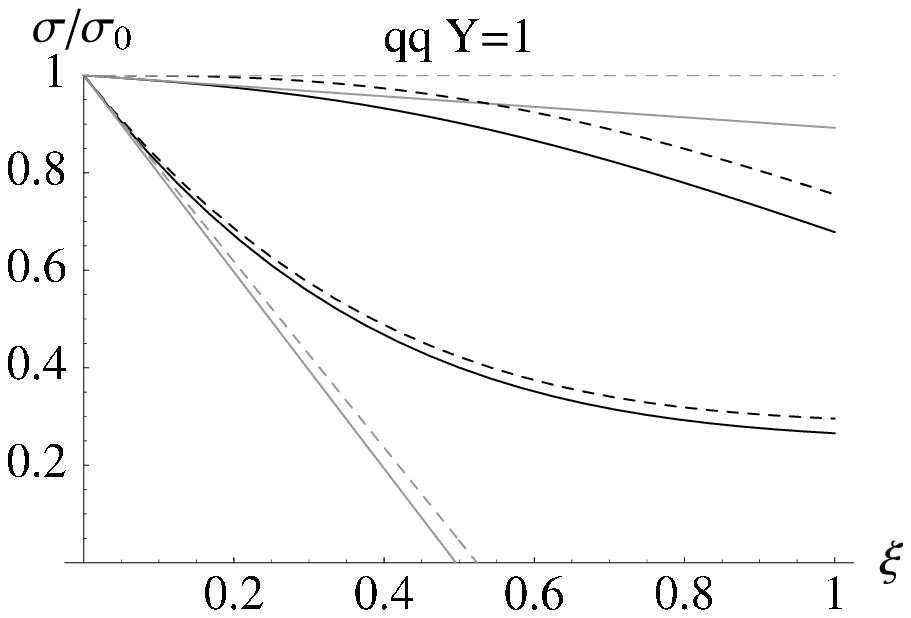,width=5cm}\epsfig{file=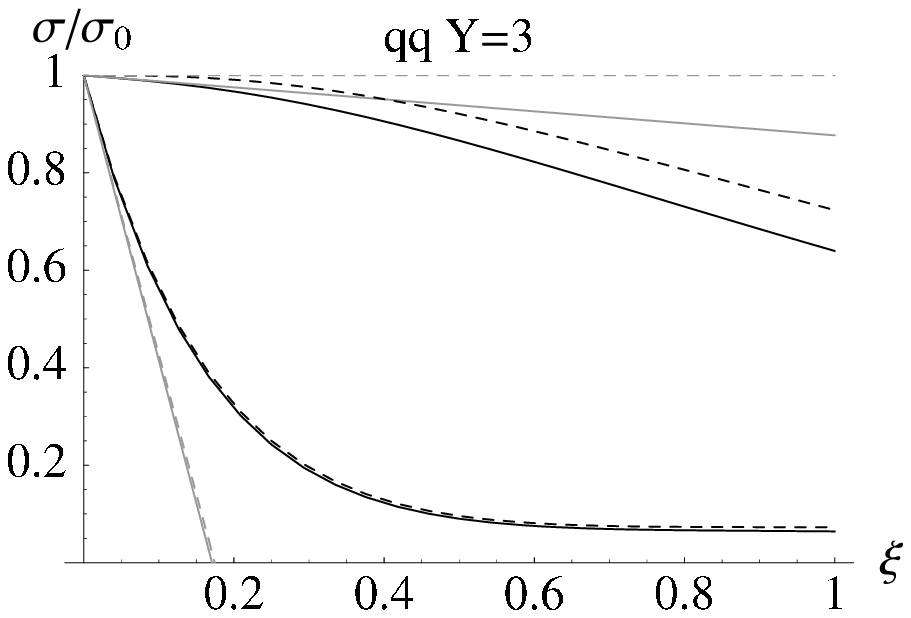,width=5cm}\epsfig{file=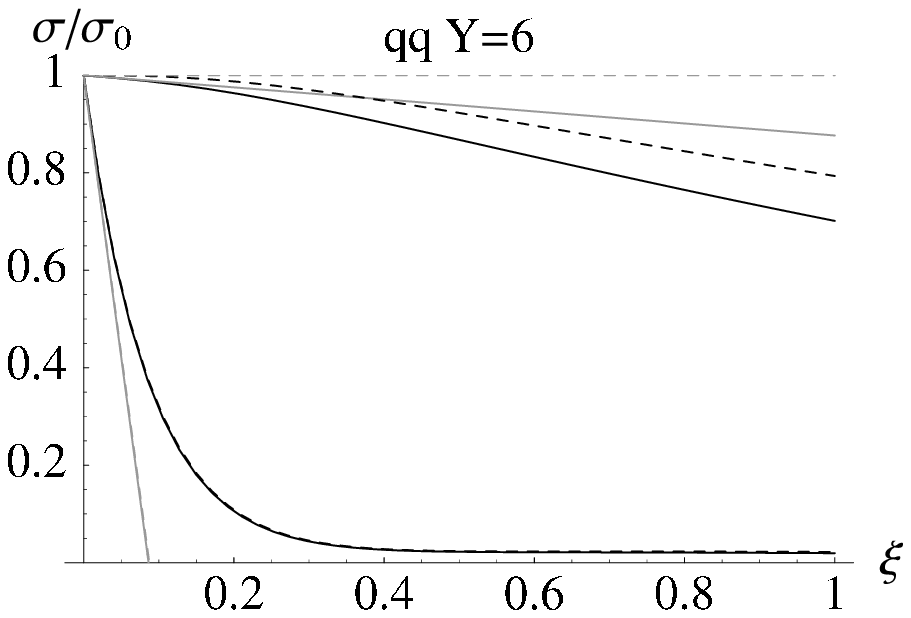,width=5cm}\\
\epsfig{file=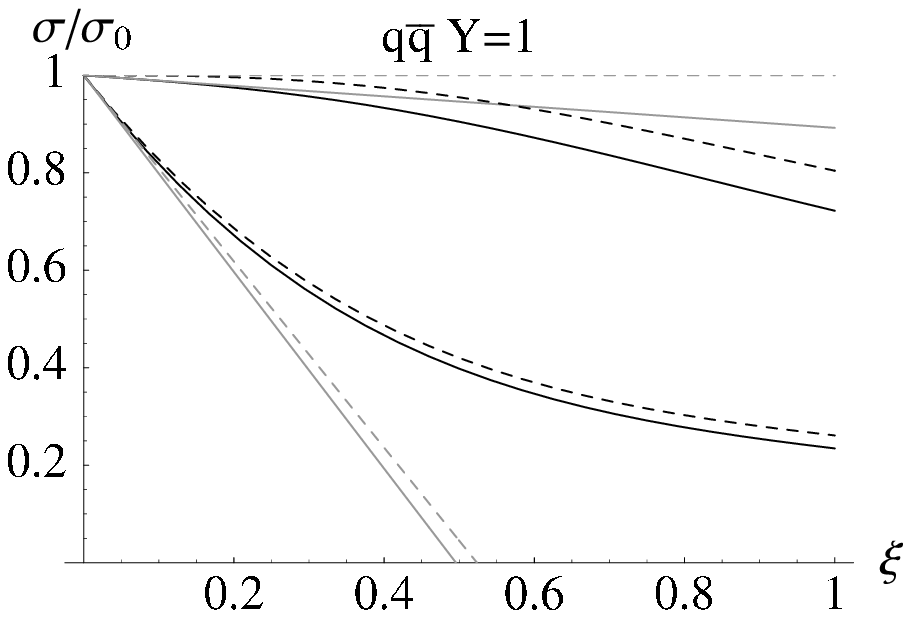,width=5cm}\epsfig{file=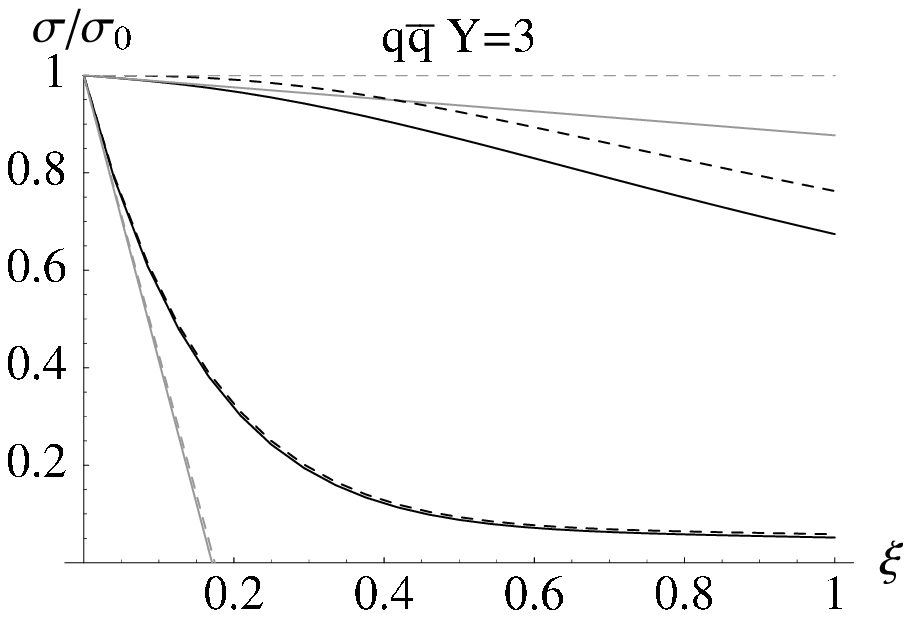,width=5cm}\epsfig{file=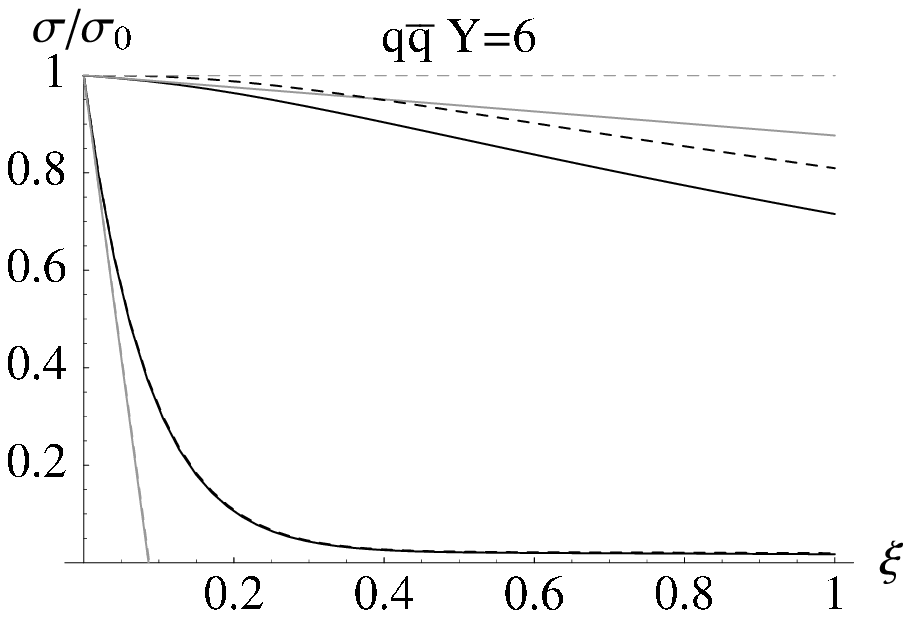,width=5cm}\\
\epsfig{file=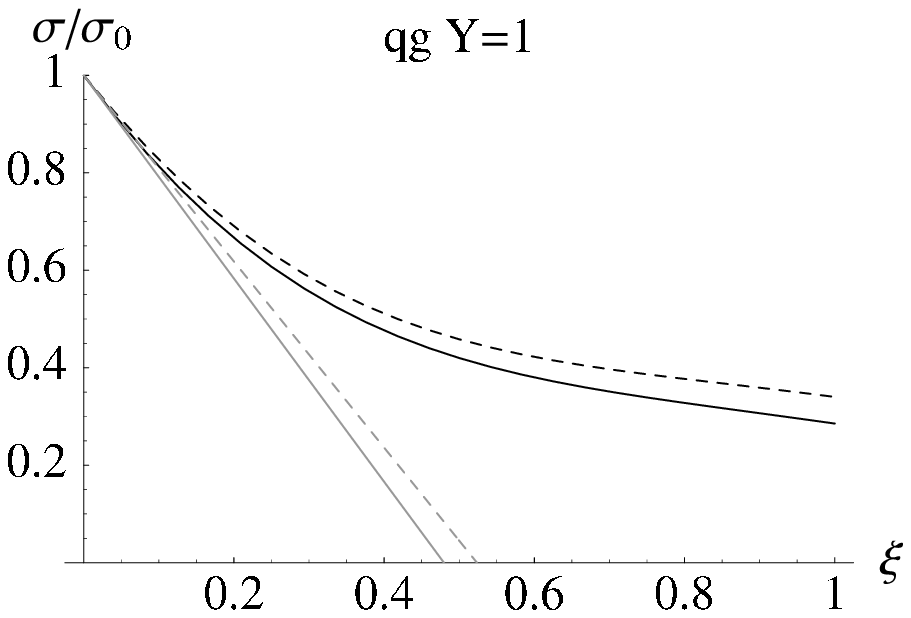,width=5cm}\epsfig{file=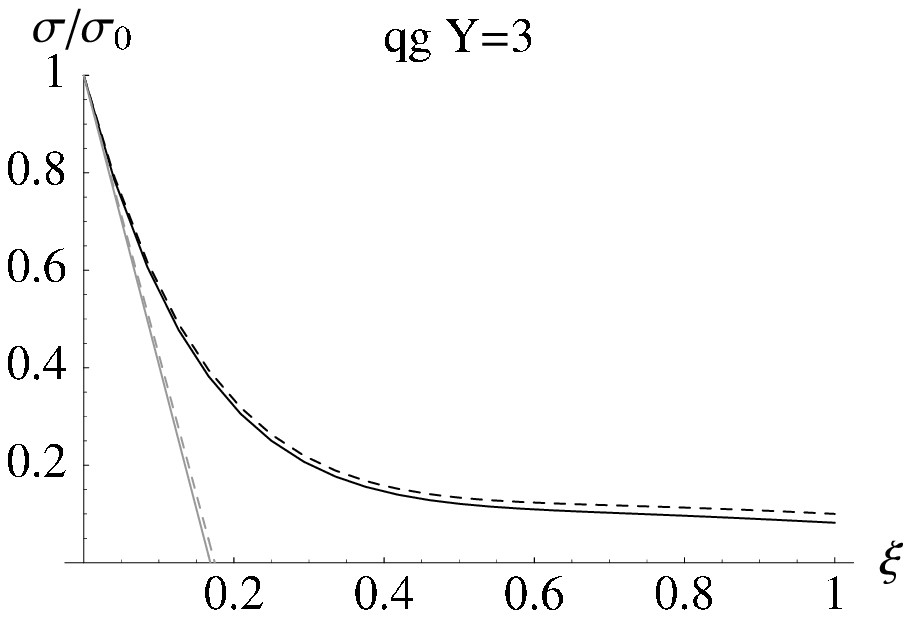,width=5cm}\epsfig{file=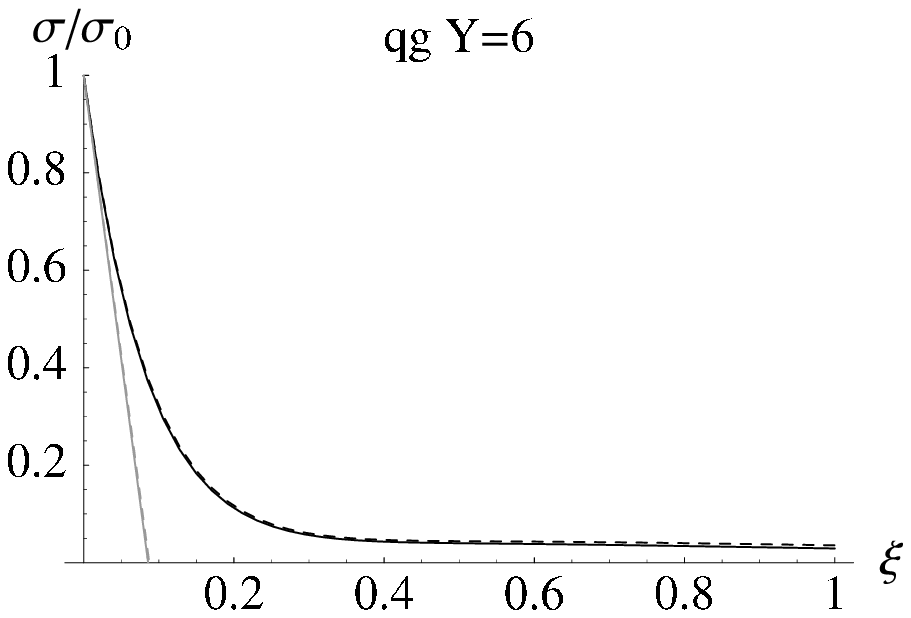,width=5cm}\\
\epsfig{file=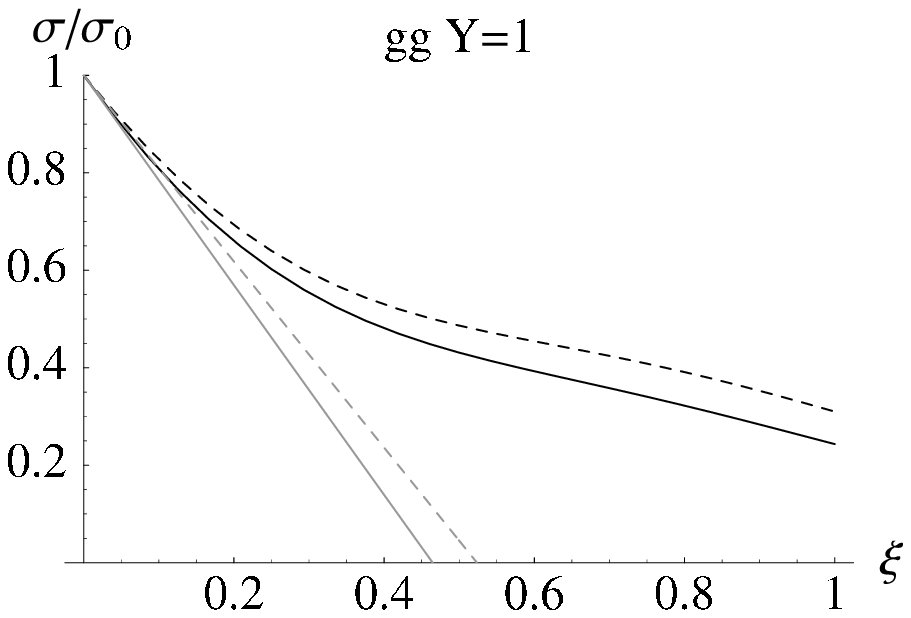,width=5cm}\epsfig{file=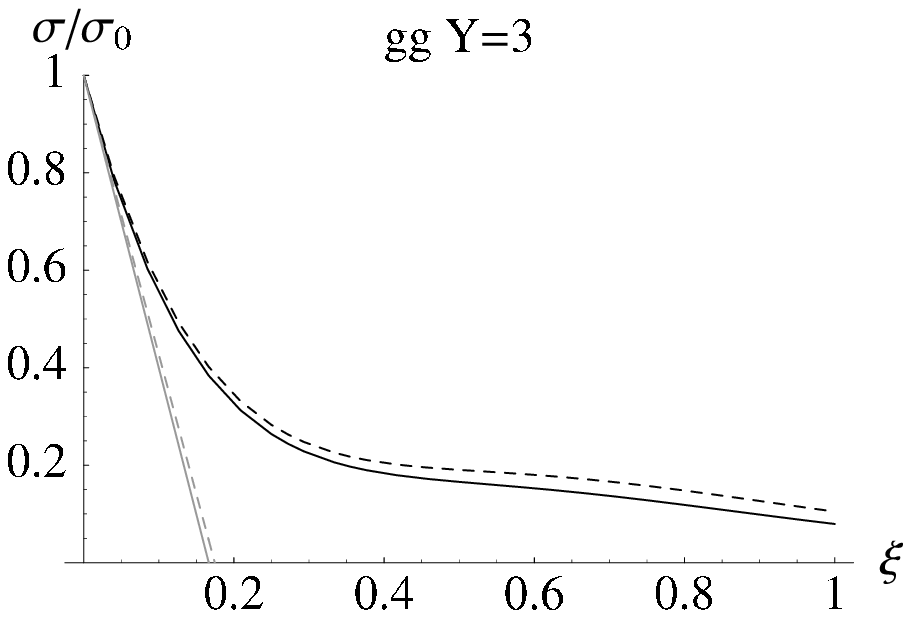,width=5cm}\epsfig{file=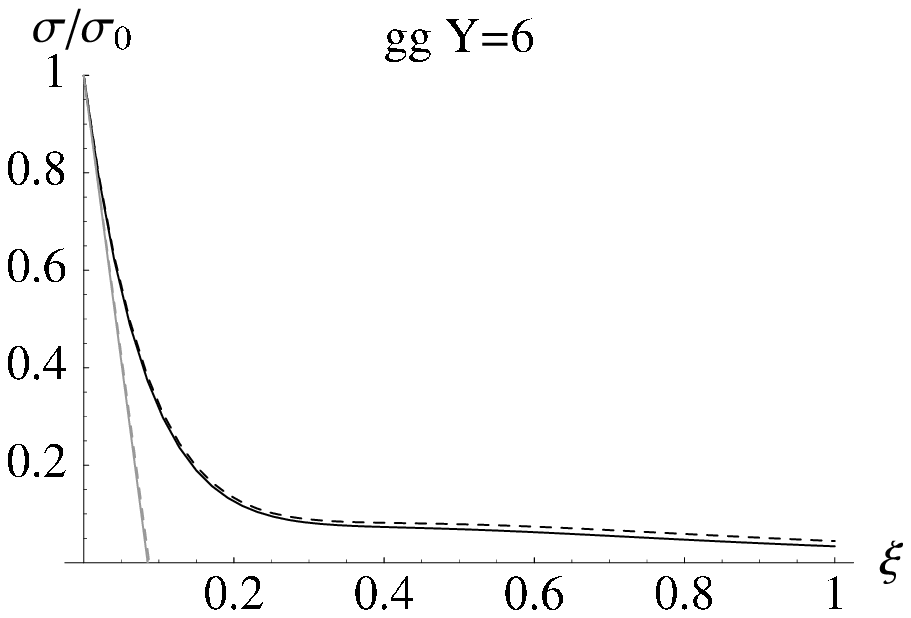,width=5cm}\\
  \caption{\label{fig:Evolution} 
    The cross-section after soft gluon evolution divided by the 
    corresponding lowest order cross-section 
    as a function of $\xi$ for $Y=1$, $Y=3$ and $Y=6$.
    The first row corresponds to quark-quark scattering, the second to quark-antiquark scattering, the third to gluon-quark
    scattering and the fourth to gluon-gluon scattering.
    The upper family of four curves in the $qq$ and $q\bar{q}$ plots are for WBF whereas all other curves are for gluon fusion.
    In all cases, the dashed curves correspond to the \pidlla and the solid lines correspond to the LLA. 
    Grey curves are the order $\as$ expansions of the corresponding black curves.}} 
 
What should also be noted is the importance of soft gluon resummation, especially for the gluon fusion process. The region around $\xi\approx 0.3$ is typical of values which will be probed at the LHC, and for such values of $\xi$ the fixed order calculation clearly is inadequate. As for the Monte Carlos, the situation is less clear since they do go beyond fixed order. Nevertheless, we know that both the soft gluon exponentiation model and the parton showering as manifest in HERWIG and PYTHIA will fail whenever the off-diagonal (Coulomb gluon) parts of $\G$ become important. The influence of the off-diagonal (i.e. octet-singlet mixing) terms is shown in \figref{fig:OS} where the grey curves are obtained by neglecting them. Their effect starts to become important for $\xi \gtaet 0.3$ (in the case of gluon fusion this value reduces slightly as $Y$ increases).

\FIGURE[t]{\epsfig{file=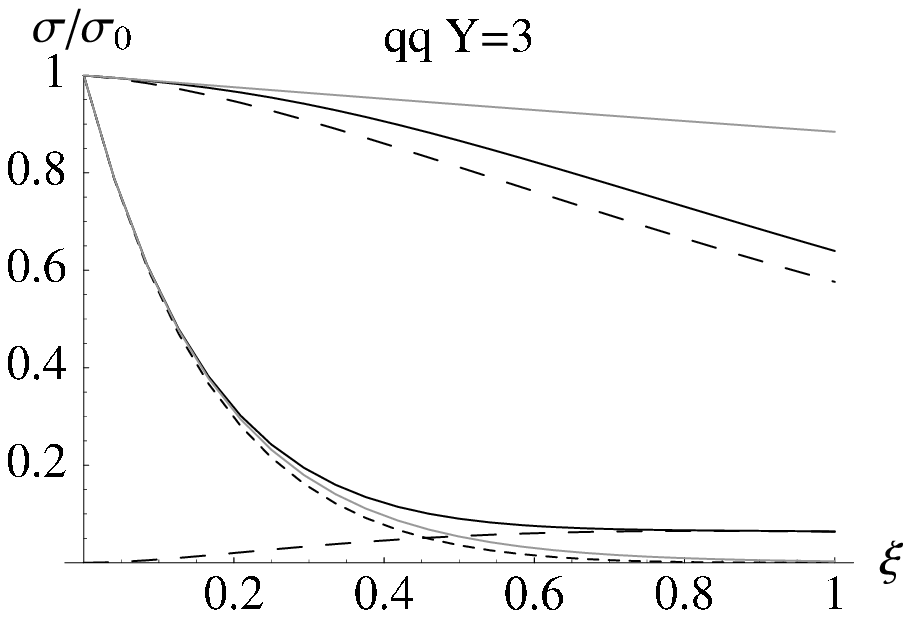,width=7.5cm}\epsfig{file=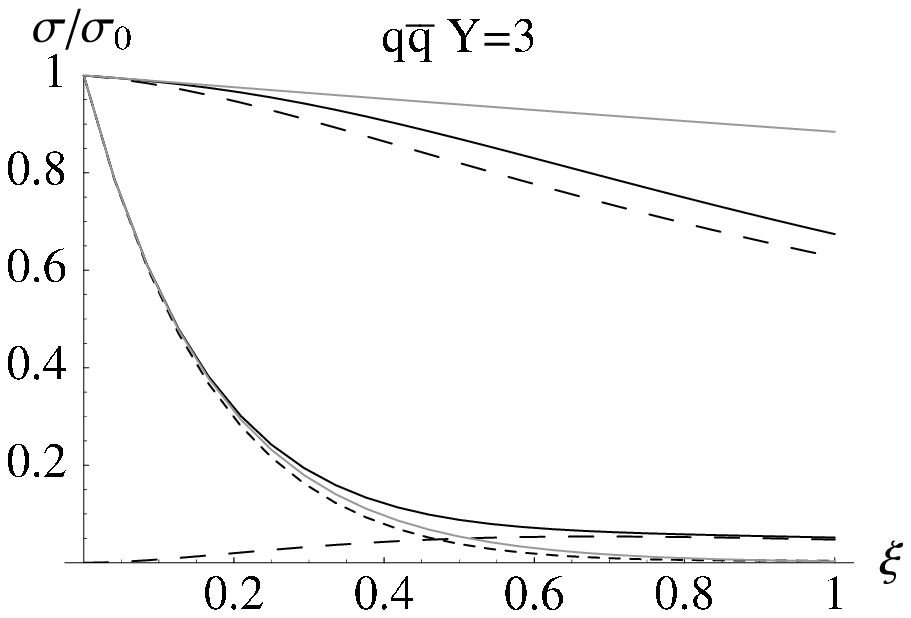,width=7.5cm}\\
\epsfig{file=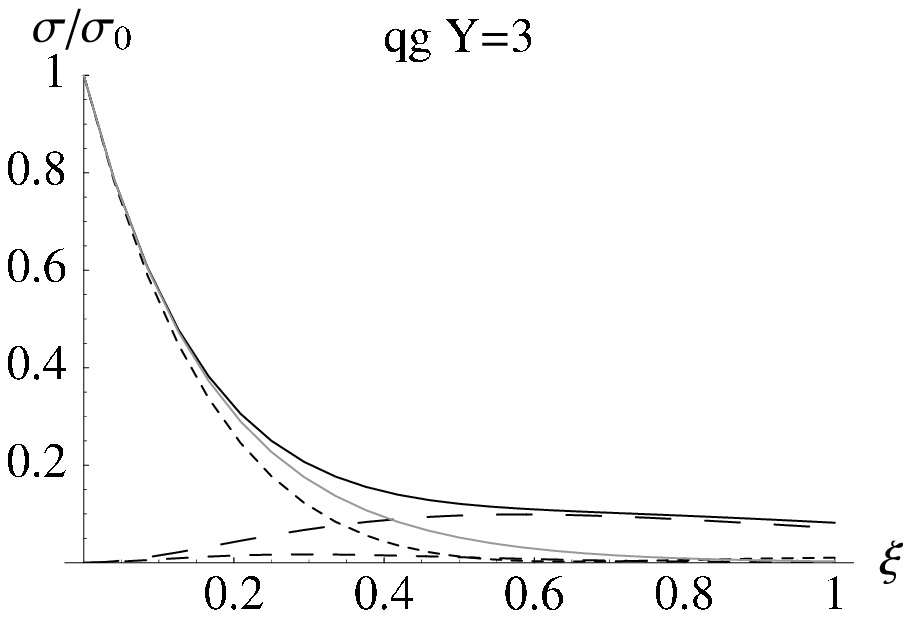,width=7.5cm}\epsfig{file=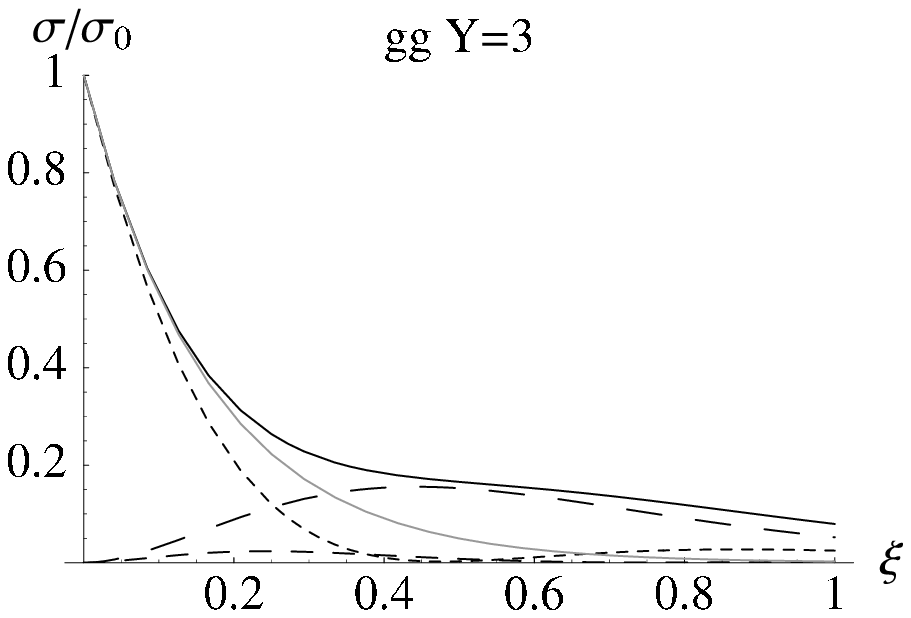,width=7.5cm}
  \caption{\label{fig:OS}  
   The cross-section after soft gluon evolution divided by the 
    corresponding lowest order cross-section 
    as a function of $\xi$ for $Y=3$.
    The solid curves are the full result (as in \figref{fig:Evolution}) and the grey curves are the result of ignoring 
    the off-diagonal entries in $\Gamma$. 
    In the $qq$ and $q\bar{q}$ cases, the upper long dashed curve is the colour singlet 
    contribution to WBF, the short dashed curve is the octet 
    contribution to the gluon fusion cross-section, the lower long dashed 
    curve is the singlet part of the gluon fusion cross-section.
    For the $qg$ and $gg$ cases, the short dashed line is again the antisymmetric octet contribution, the long dashed line
    is the singlet contribution whilst the medium dashed line corresponds to the small symmetric octet contribution. The
    10 and 27 exchange contributions formally present in the $gg$ case are too small to be visible.}}
 
The colour mixing is related to another striking feature of  \figref{fig:Evolution}, namely the fact that the evolution of the gluon fusion contribution starts out as an exponential decay, but then evolves to a much 
slower decay. The exponential decay for low $\xi$ is expected since both the $\rho$ and the $Y$ terms in $\G$ represent  exponential decays in  $\xi$. The non-exponential property is thus a consequence of the mixing induced by the off-diagonal imaginary parts of the evolution matrix. This is clarified in \figref{fig:OS} where the different colour components are shown separately. In fact this behaviour is arising because a part of the perturbative QCD pomeron is present in our calculations and it is 
visible in \figref{fig:OS} as the singlet part of the initial octet exchange (long dashed line).  For sufficiently large $Y$ the resummation we perform here will not be adequate and the BFKL resummation of the series in $\as Y$ will become important \cite{Kuraev:1977fs,Balitsky:1978ic,Mueller:1992pe}. The problem of summing simultaneously the soft and high-energy logarithms has been explored in \cite{Forshaw:2005sx} where it was illustrated that BFKL effects may well be important for $Y$ values as small as 3. 
In the $qq$ and $q\bar{q}$ cases, we note that the long dashed curve represents not only the singlet contribution to the gluon fusion cross-section but also the octet contribution to WBF. That these two are identical in the leading logarithmic approximation follows from the symmetry of the evolution matrix in a basis where ${\mathbf S}=\mathrm{diag}(1,1)$ \cite{Seymour:2005ze}.

Finally we turn our attention to the interference between the $ZZ$ fusion and gluon-gluon fusion contributions. In our calculations there is no interference if one integrates over the $\cos \phi$ dependence of the tag jets since soft gluons do not change the original azimuthal dependence of the lowest order matrix elements. A sizeable interference would certainly be
of interest since the azimuthal angular distribution $\sim \cos \phi$ differs both from the flat WBF distribution and
the gluon-gluon fusion distribution $\sim \cos^2\phi$. The three different angular distributions would in principle provide a handle which would help isolate the separate contributions. 

However, we find that the level of interference is very low. In fact, even if the color structure is neglected, there is a cancellation between quarks of different helicity and flavour already at tree level. In the approximation of vanishing $\sin^2 \theta_W$ and assuming only u-type quarks, the level of cancellation due to the helicity structure is 1/4, since in only one case out of four will there be two left handed quarks interacting. This reduces to 1/36 if we assume that the proton is made up of two up quarks and one down quark which in turn reduces dramatically to $1.6 \times 10^{-4}$ when we restore the correct value of the weak mixing angle.\footnote{In fact the interference would have vanished identically if $\sin^2\theta_W = 1/4$.}

Writing the interference between $ZZ$ and $gg$ fusion contributions as
\begin{eqnarray}
  I &=& \M^{\dagger}\Sv \M
  -\M^{\dagger}_{gg}\Sv \M_{gg}
  -\M^{\dagger}_{ZZ}\Sv \M_{ZZ},
\end{eqnarray}
where $\M=\M_{gg}+\M_{ZZ}$ is given in \eqref{eq:M},  we show $-I/\sigma_0$ in \figref{fig:Mix} for the case of u quark scattering: as anticipated it is very small.
				   
\FIGURE[h]{\epsfig{file=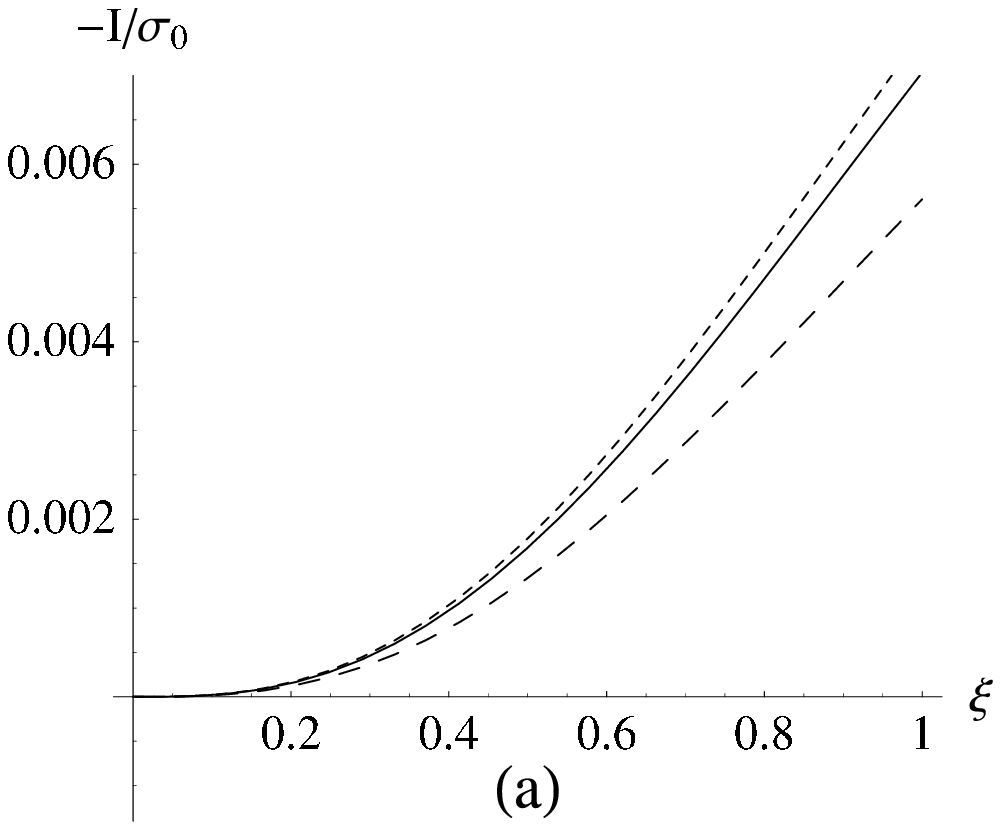,width=5cm}\epsfig{file=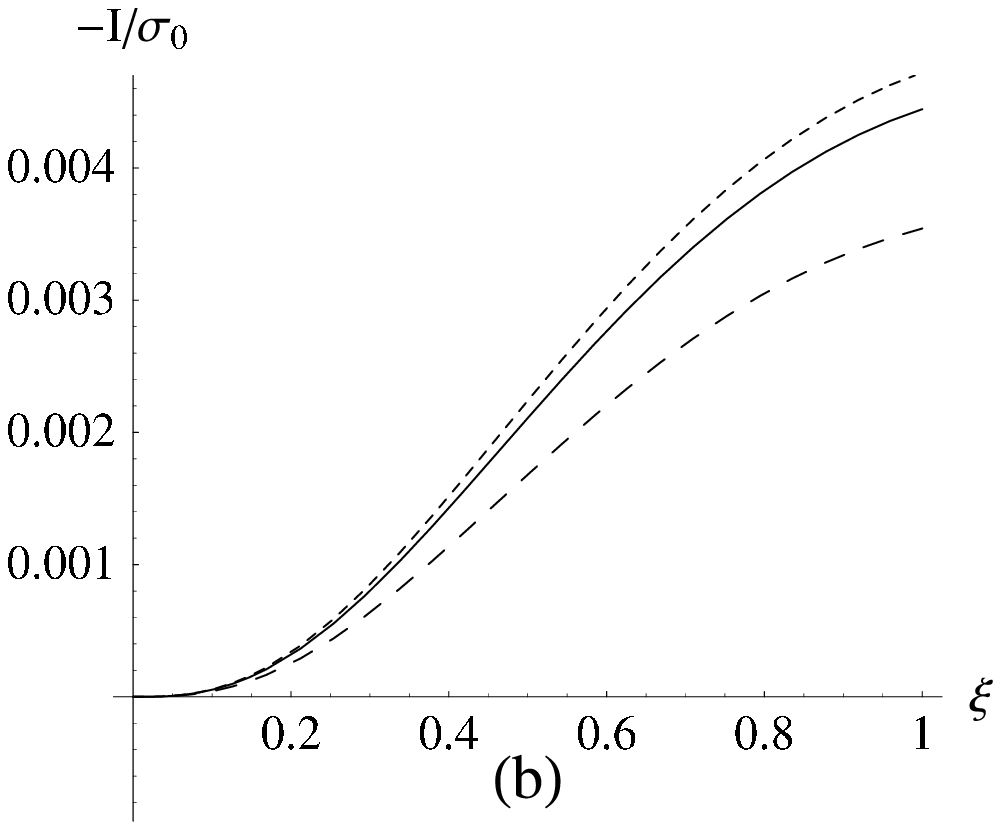,width=5cm}\epsfig{
    file=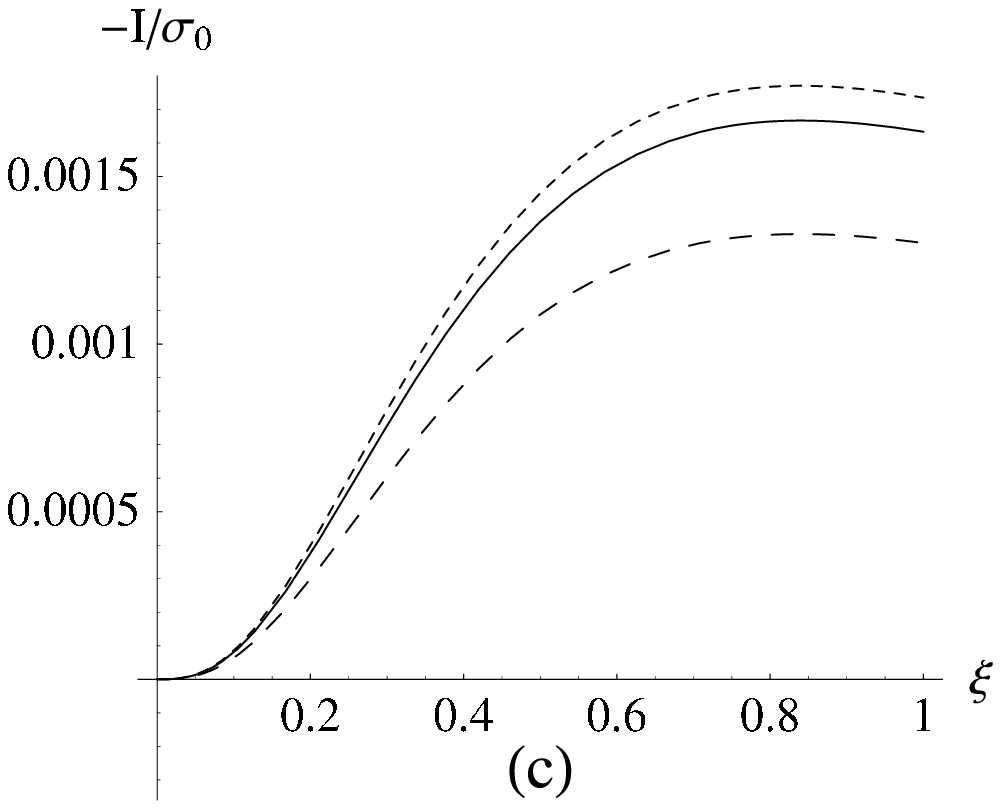,width=5cm}
  \caption{\label{fig:Mix} The size of the $ggH$-$ZZH$ interference
    (normalized to the total lowest order cross-section for $gg$ and $ZZ$ fusion)
    in the case of u-type quarks only. 
    In (a) $Y=1$, (b) $Y=3$, (c) $Y=6$. Solid curves are for $|\kT|=30$, 
    long dashed curves for $|\kT|=100$ and short dashed curves for $|\kT|=300$.
}}

\section{Conclusion}
\label{sec:Conclusions}

We have performed a partial summation of the leading soft gluon contributions in Higgs plus two jet production in the presence of a (mini-)jet veto. The imposition of such a veto helps enhance the contribution from weak boson fusion by exploiting the fact that no colour is exchanged
between the outgoing tag jets (and hence radiation is suppressed in the central region). We find that large differences in (mini-)jet activity can indeed be expected depending on whether the hard process is a color octet gluon fusion process or a color singlet weak boson fusion process. 

Neglecting non-eikonal contributions (which is a good approximation when the tag jets are at low angles relative to the collision axis) the probability for extra gluon emission in the gluon fusion case is proportional to $\as N Y$ and hence increases as the gap between the tag jets increases. For the electroweak case it is proportional to $\as C_F \rho$ which implies that radiation is not strongly enhanced as the gap increases. This is apparent in the \pidlla where one keeps only the leading $Y$ and $i\pi$ dependent terms. In that case, corrections to the WBF process starts at ${\cal O}(\as^3)$.
Apart from containing the expected general features, our results also indicate when it is necessary to go beyond fixed order perturbation theory and resum the logarithms in the ratio of the tag scale to the veto scale. We conclude that for both the gluon fusion and weak boson fusion production of a Higgs, resummation is important in the region which can be probed at the LHC. Moreover, our results indicate that for large enough values of the evolution variable, $\xi$, it is not sufficient to account for the resummation using existing parton shower algorithms or the soft gluon exponentiation model.

We have also made an estimate of the size of the interference contribution between weak boson fusion and gluon-gluon fusion.  The result is very small, never exceeding 1\% of the lowest order cross-section.

\section*{Acknowledgments}

We thank Mrinal Dasgupta, Lance Dixon, Leif L\"onnblad, Mike Seymour and Torbj\"orn Sj\"ostrand for valuable discussions. This project was supported in part by the UK Science and Technology Facilities Council.

\appendix
\section{Evolution matrices}  
\label{AppA}
In this appendix we list the evolution matrices for the various subprocesses relevant to Higgs production through
gluon fusion. The function $\rho(Y,y_3,y_4)$ is as defined in Eq.(\ref{eq:rho}) and this is the only difference compared
to the evolution matrices presented in \cite{Kidonakis:1998nf,Oderda:1999kr,Appleby:2003sj}.

For $q\bar{q} \to q\bar{q}H$ the evolution proceeds in a similar fashion to the $qq$ case except that the evolution matrix, in the singlet-octet basis, is
\begin{equation}
  \G_{q\bar{q}} = \frac{C_F}{2}\rho(Y,y_3,y_4) \mathbf{1}+
  \left(
  \begin{array}
    [c]{cc}
    0  & -\frac{N^{2}-1}{4N^{2}}i\pi\\
    -i\pi & (\frac{N}{2}Y-\frac{N^2-2}{2N}i\pi)
  \end{array}
  \right).
\end{equation}
For $qg \to qgH$ we use the basis
\beqa
c_1&=&\delta_{\e_A\e_1}\delta_{\e_B\e_2}\nonumber\\
c_2&=&d_{\e_B \e_2 c}{\left( T^c \right)}_{\e_1 \e_A}\nonumber\\
c_3&=&if_{\e_B \e_2 c}{\left( T^c \right)}_{\e_1 \e_A} \, 
\label{qgqgbasis}
\eeqa 
corresponding to singlet, symmetric octet and antisymmetric octet $t$-channel exchange.
In this case $\mathbf{S}_{qg}$ is
\begin{equation}
\mathbf{S}_{qg} =
\left( 
\begin{array}{cccll|}
  N(N^2-1)  & 0 & 0 \\
  0 & \frac{1}{2N}(N^2-4)(N^2-1) &0\\
  0 & 0 &\frac{1}{2} N(N^2-1)  
\end{array}
\right)
\label{eq:Sqg}
\end{equation}
and the evolution matrix is
\begin{equation}
  \mathbf{\Gamma}_{qg} = \frac{(3N^2-1)}{8N} \rho(Y,y_3,y_4) \; \mathbf{1} + \frac{1}{4}
\left(
\begin{array}{ccclll}
 N \pi i & 0 & 2 \pi i  \\
 0 & 2NY & N \pi i  \\
 4 \pi i  & \frac{\left(N^2-4\right)}{N}\pi i & 2NY \\
\end{array}
\right)\; .
\label{qgqgano}
\end{equation}
For $gg \to ggH$ we use the basis (setting $N=3$)
\beqa
C_1&=&\frac{1}{8}\delta_{\e_A \e_1} \delta_{\e_B \e_2} \, ,
\nonumber \\ 
C_2&=&\frac{3}{5} d_{\e_A\e_1c} d_{\e_B\e_2c} \, ,
\nonumber \\ 
C_3&=&\frac{1}{3} f_{\e_A\e_1c} f_{\e_B\e_2c} \, ,
\nonumber \\ 
C_4&=&
\frac{1}{2}(\delta_{\e_A \e_B} \delta_{\e_1 \e_2}
-\delta_{\e_A \e_2} \delta_{\e_B \e_1}) -\frac{1}{3} f_{\e_A\e_1c} f_{\e_B\e_2c} \, ,
\nonumber \\ 
C_5&=&
\frac{1}{2}(\delta_{\e_A \e_B} \delta_{\e_1 \e_2}
+\delta_{\e_A \e_2} \delta_{\e_B \e_1})-\frac{1}{8}\delta_{\e_A \e_1} \delta_{\e_B \e_2}
-\frac{3}{5} d_{\e_A\e_1c} d_{\e_B\e_2c} \, .
\label{ggggbasis}
\eeqa 
In which case 
\beq
\mathbf{S}_{gg}=\left( 
\begin{array}{ccccc}
1 & 0 &0 & 0 & 0      \\  
0 & 8 &0 & 0 & 0      \\
0 & 0 &8 &0 &0      \\
0 & 0 &0 &20 &0     \\
0 & 0 &0 &0 &27     
\end{array}
\right) \, 
\label{ggggsoft}
\eeq
and
\beqa
{\Gamma}_{gg}=\
\frac{3}{2} \rho(Y,y_3,y_4) \; \mathbf{1} +
\left(
\begin{array}{ccccclllll}
 -\frac{3}{4}\pi i & 0 & -3 \pi i & 0 & 0 \\
 0 & \frac{3}{2}Y & -\frac{3}{4} \pi i & -\frac{3}{2} \pi i & 0 \\
 -\frac{3}{8} \pi i & - \frac{3}{4} \pi i & \frac{3}{2} Y & 0 & -\frac{9}{8} \pi i\\
 0 & -\frac{3}{5} \pi i & 0 & 3Y - \frac{3}{4} \pi i & -\frac{9}{10}\pi i \\
 0 & 0 & -\frac{1}{3} \pi i & -\frac{2}{3} \pi i & 4Y - \frac{5}{4} \pi i
\end{array}
\right).
\eeqa  

\bibliographystyle{utcaps}  
\bibliography{refs,Hrefs} 

\end{document}